HS-45

# Pan-Cancer Epigenetic Biomarker Selection from Blood Samples Using SAS®

Xi Chen, University of Kentucky; Jin Xie, University of Kentucky; Qingcong Yuan, Miami University

## ABSTRACT

A key focus in current cancer research is the discovery of cancer biomarkers that allow earlier detection with high accuracy and lower costs for both patients and hospitals. Blood samples have long been used as a health status indicator, but DNA methylation signatures in blood have not been fully appreciated in cancer research. Historically, analysis of cancer has been conducted directly with the patient's tumor or related tissues. Such analyses allow physicians to diagnose a patient's health and cancer status; however, physicians must observe certain symptoms that prompt them to use biopsies or imaging to verify the diagnosis. This is a post-hoc approach. Our study will focus on epigenetic information for cancer detection, specifically information about DNA methylation in human peripheral blood samples in cancer discordant monozygotic twin-pairs. This information might be able to help us detect cancer much earlier, before the first symptom appears. Several other types of epigenetic data can also be used, but here we demonstrate the potential of blood DNA methylation data as a biomarker for pan-cancer using SAS® 9.3 and SAS® EM. We report that 55 methylation CpG sites measurable in blood samples can be used as biomarkers for early cancer detection and classification.

## INTRODUCTION

Cancer has slowly replaced heart disease and become the number one leading cause of death in the United States as of early 2014. The world-wide cancer incidence rate is still increasing; thus, early detection through the use of biomarker analysis is a key focus in current cancer research. Two criteria of interest are performance—sensitivity, specificity and predictive value, and accessibility—that is, low cost and easy access without complicated surgeries (Hartwell et al. 2006). Blood is one of the most abundant tissues in the human body, and the process of obtaining blood samples is simple and routine. Blood samples have long been used as a health status indicator, but DNA methylation signatures in blood have not been fully appreciated in cancer research.

Cancer diagnoses have typically involved analyses conducted directly on tumor or otherwise cancerous tissue. Thus, to determine a patient's cancer status, physicians have needed to observe certain symptoms, then use either biopsy or imaging to verify the diagnosis. If blood samples can be used instead of these more invasive techniques, we may be able to detect cancer much earlier, potentially even before the first symptom appears. Such an analysis may be possible by examining epigenetic features (patterns) present in blood.

Waddington coined the term "epigenetics" in the 1950's. He defined it as changes in phenotype (cellular level), without changes in genotype (DNA level). We now understand that epigenetic mechanisms manipulate gene expression patterns without altering the DNA sequence through several known modes, including DNA methylation, histone modification, and chromatin remodeling (Stunnenberg and Hirst 2016). In the work described here, we have focused on DNA methylation.

Our understanding of how DNA methylation changes during cellular differentiation, normal development and how it relates to other epigenetic mechanisms remains limited. Methylation can only happen on one of the four DNA bases, cytosine, to form 5-methylcytosine. Methylation often occurs at CpG sites, where a cytosine nucleotide is followed by a guanine nucleotide in the 5' to 3' direction(Baylin 2005). CpG islands are regions with a high frequency of CpG sites, typically 300–3000 base pairs in length for humans. About 70% of human promoter regions (proximal promoters) have a high CpG content (Deaton and Bird 2011), thus methylation of these regions plays an important role in regulation of gene expression. Methylation signatures may differ among tissues and between normal and abnormal cells within the same individual.

Aberrant methylation patterns are associated with a number of diseases, age, and environmental factors such as smoking or drug abuse (Bjornsson 2008). Therefore, methylation status is a potential candidate as a cancer biomarker. Epigenetic sequencing could potentially provide more direct information on disease status; however, the data from this method share the same notorious properties as other genomic methods: multimodality and high dimensionality (Laird, 2010). More often than not, only a small portion of these biomarker features are studied and utilized for disease detection.

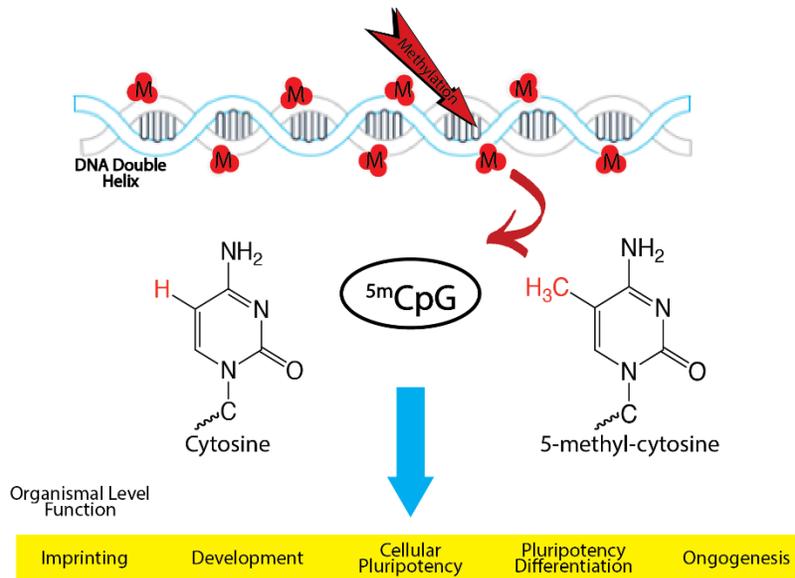

**Figure 1. Methylation at the cellular level. DNA methylation is essential for normal development and has been implicated in many pathologies, including cancers. One of the 4 bases in DNA, cytosine, undergoes methylation at its carbon-5 position. At the organism level, methylation play important roles in imprinting, development, cellular pluripotency, pluripotency differentiation, oncogenesis, etc.**

Many biological features are candidates for use as biomarkers, but the key aspects characterizing ideal biomarkers are high accuracy, low rate of false negatives, and easy accessibility. Blood is a tissue type that is a good source of biomarkers. Blood is one of the most abundant and accessible tissues in the human body, and the methylation signatures (epigenetic information) found in blood samples provide rich information and demonstrate the capability to differentiate cancer status.

DNA methylation signatures in blood have been associated with breast, colon, bladder, and ovarian cancers. To the best of our knowledge, only a limited number of studies have attempted to identify epigenetic biomarkers associated with pan-cancer (cancer of multiple organs), and none have approached this problem from a statistical perspective(Esteller 2009). Through identifying malignancy-associated DNA methylation changes in blood, we propose a statistical approach/pipeline to explore the potential of blood DNA methylation as a biomarker for pan-cancer.

## METHODS

### DATASETS

Epigenome-wide DNA methylation profile data were curated by the International Human Epigenome Consortium (IHEC) and are publicly accessible from Gene Expression Omnibus (GEO) (Barrett et al. 2013). IHEC allows us to easily select epigenetic data based on consortium, tissue, and assay type. The GEO accession code for the data used for this study is GSE89093; this dataset consists of 92 methylation profiles from the HumanMethylation450 BeadChip (450k) platform. Each profile comprises 453,627 methylation sites. The methylation profile provides a functional assessment by evaluating the plasma levels of methionine, cysteine, SAM, SAH, homocysteine and cystathionine. The dataset has been normalized at single CpG resolution as beta values with a numerical range between 0 (unmethylated) and 1 (methylated); the full dataset can be accessed at GEO website.[1] Beta values were obtained from the ratio of signal intensity from the methylated probes over the sum of signal intensity from both unmethylated and methylated probes (Roos et al. 2016).

## DATA COLLECTION

Genome-wide peripheral blood DNA methylation profiling (study data) samples were collected from the National Cancer Registry at the Office for National Statistics (ONS) for twin-pairs registered with the TwinsUK Adult Twin Registry. The participants were 46 adult female monozygotic twin-pairs; samples obtained at the same time point, from 41 healthy female participants (not diagnosed with any cancer) paired with their monozygotic (MZ) co-twin who were diagnosed with cancer within a five-year window. Five extra pairs with a co-twin diagnosed with cancer 5 to 11 years prior to the blood sampling time point were also included. A total of 8 types of cancers were included this study; their distribution is included in Figure 2.

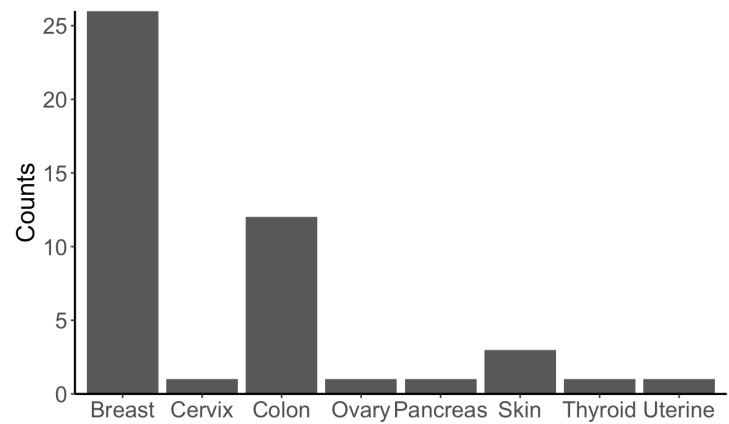

**Figure 2. Cancer typing distribution. Eight common types of cancer were present among 41 participants: breast, cervix, colon, ovary, pancreas, skin, thyroid, and uterine cancers. However, due to the limited sample size, the cancer types are not evenly distributed, with most cases breast and colon cancers.**

## DATA CLEANING/VALIDATION

To analyze the data with SAS® and SAS® EM, the raw data needed to be reformatted. The feature dimension of the data was rather large: 453,627 features (columns) from 92 observations (rows). First, the responsive variable column was concatenated to the data. Information from the metadata accompanied the original data and indicated the disease status; cancer or healthy. We also included data cleaning steps, including transformation.

To validate the data, we used SAS® PROC UNIVARIATE to generate statistics: (1) moments, (2) basic measures of location and variability, (3) tests for location, (4) quantiles, (5) extreme observations. The data were then ready for the next step of analysis.

## TRAINING-TESTING DATA SPLITING

Since the goal of this analysis was to identify potential biomarkers, narrowing 453,627 candidates to only a small number of methylated CpG sites, we applied standard machine learning approaches. As for any statistical analysis, we partitioned the data by randomly selecting 70% (64 observations) of the individuals as the training set, and the remaining 30% (28 observations) as the testing set. Aiming to be representative, we maintained the cancer and healthy profile ratio of the original data in these training and testing sets.

## VARIABLE SCREENING

---

[1] GEO dataset link: https://www.ncbi.nlm.nih.gov/geo/query/acc.cgi?acc=GSE89093.

Several methods were used to reduce the number of variables to a moderate range. We first computed the association between each CpG site and the response variable, then ranked the association from highest to lowest based on importance. This step includes four methods of variable screening: Pearson correlation, T-test, distance correlation, and expectation of conditional difference (ECD). From each method, we selected the 100 CpG sites with the highest ranking as potential candidates, and finally reported a subset of 55 common CpG sites across methods as our final choices (Figure 3).

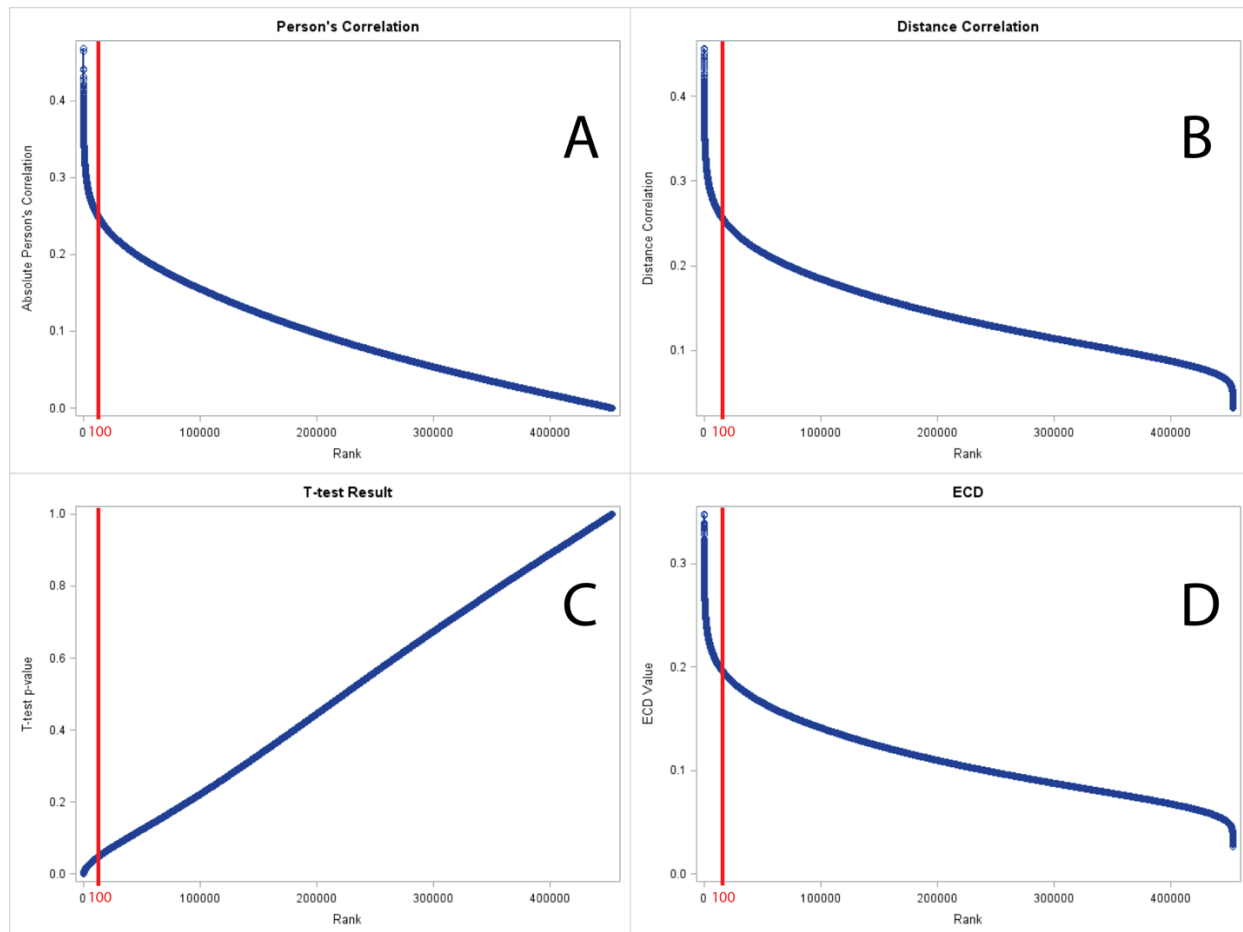

**Figure 3. Variable Screen step. A: Sure Independence Screening; B: Independence Screening via Distance Correlation; C: T-test Analysis; D: Expectation of the Conditional Difference. Red line in the figure shows the first 100 variables according to the ranking of each approach.**

**Sure Independence Screening.** With the advent of modern technology for data collection, researchers are able to collect ultra-high dimensional data at relatively low cost in diverse fields of scientific research, such DNA methylation data. (Fan and Lv 2008) proposed the Sure Independence Screening (SIS) algorithm and shown that the Pearson correlation ranking procedure possesses a sure screening property. That is, all truly important predictors can be selected with a probability approaching one as the sample size diverges to infinity. We calculated the absolute value of the Pearson correlation between each predictor and response variable, as shown in Figure 4. We ordered the absolute values of Pearson correlation from high to low and picked the top 100 CpG sites (Figure 3A,).

**Independence screening via Distance Correlation.** (Székely, Rizzo, and Bakirov 2007) proposed distance correlation and showed that the distance correlation of two random vectors equals zero if and only if these two random vectors are independent. Furthermore, the distance correlation of two univariate normal random variables is a strictly increasing function of the absolute value of the Pearson correlation of these two normal random variables. Li, Zhong and Zhu (2012) developed a sure independence screening procedure based on the distance correlation (DC-SIS), which is a model-free variable screening algorithm that enjoys the sure screening property. We used it to calculate the distance correlation between each CpG site and the response variable. As shown in Figure 3B, we also ranked the distance correlation values from high to low and picked the top 100 CpG sites.

**T-test analysis.** A T-test has long been considered the most direct analysis approach (Winer, Brown, and Michels 1971). For this study, we conducted an independent t-test to compare DNA methylation profiling in the cancer and healthy groups. CpG sites with smaller p-values indicate a higher difference between the two groups. We report the p-value and rank from lowest to highest in Figure 3C.

**ECD analysis.** The expectation of conditional difference (ECD) is an independence measure developed by (Yin and Yuan, n.d.) that measures the association between two sets of random variables, especially when one of them is categorical or discrete. The measure is a value between 0 and 1; a higher value indicates a stronger association. We recorded the ECD measures and ranked them from highest to lowest, as shown in Figure 3D.

## MACHINE LEARNING ALGORITHMS

We also implemented several machine learning algorithms using SAS® EM. These algorithms include Random Forest, Neural Network and Support Vector Machine (SVM), and their implementation pipeline is depicted in Figure 4.

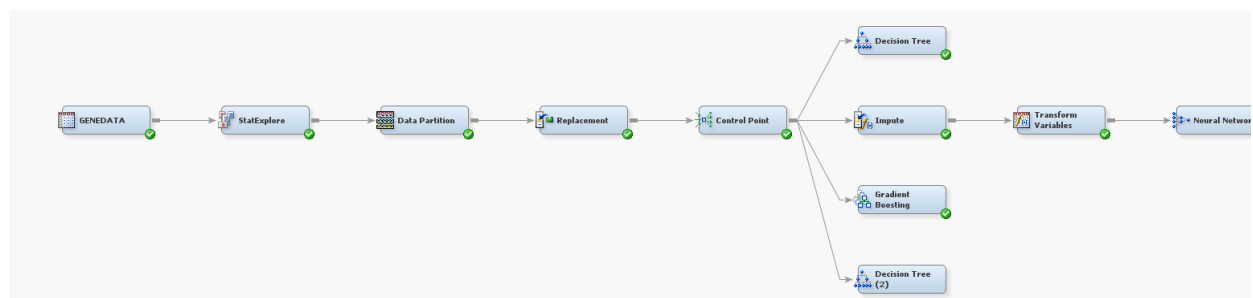

**Figure 4. SAS® EM-generated implementation pipeline for Random Forest, Neural Network and Support Vector Machine algorithms.**

## RESULTS

After the variable screening step, we identified 55 common CpG sites among the top 100 candidates identified by each of the four different methods described above. We incorporated these 55 common CpG sites into our predictive model via Linear Discriminant Analysis of the training set. Model performance is reported in Table 1. Our algorithm reported a high accuracy of 71.4% on the testing set, and a false negative rate (FNR) of 14.3%. The false positive rate (FPR) is 42.9%; we believe this rate is acceptable, since the FNR is more important in clinical settings (Norris and Kahn, 2006). Our algorithm also reported a receiver operating characteristic (ROC) curve (Figure 4); the area under the curve (AUC) is 0.658.

| Data Set | Accuracy | FPR | FNR |
|---|---|---|---|
| Training Set (70%) | 100% | 0% | 0% |
| Testing Set (30%) | 71.4% | 42.9% | 14.3% |

**Table 1 Model performance (evaluation) summary.**

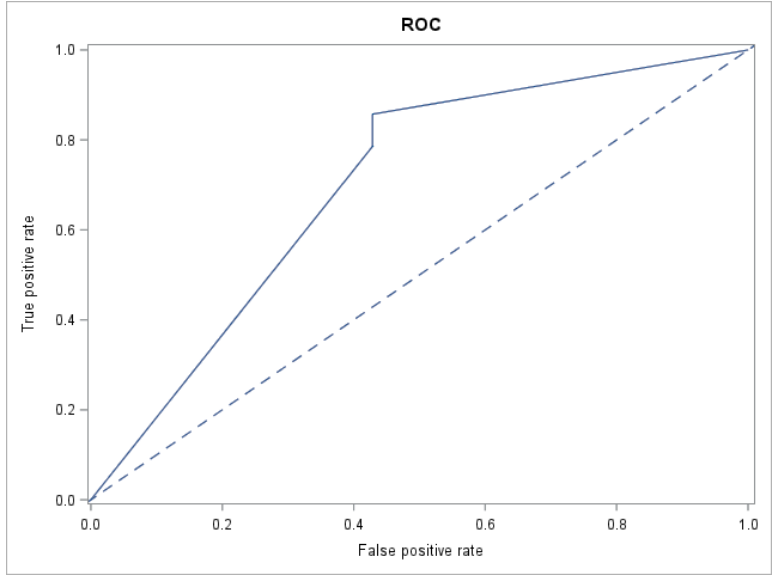

**Figure 4. Receiver operating characteristic (ROC) curve generated using linear discriminate analysis (LDA) for the testing dataset examining the 55 common CpG sites selected from the variable screening step.**

We also applied several different classification algorithms such as Random Forest, Neural Network and Support Vector Machine (SVM) provided from SAS® EM. Unfortunately, the performance of these algorithms was not desirable. With or without a variable screening step, these algorithms had an accuracy around 50%, the same rate expected for random guessing. Therefore, our simple statistical model algorithm outperformed these machine learning algorithms.

To help other researchers understand our results, we decided to map the epigenetic CpG sites back to the genetic data. Since cancerous and healthy cells will have different methylation patterns (Feinberg & Vogelstein, 1983), important (high-ranked) CpG sites should correlate with the importance of genes. For the 55 CpG sites we identified 55 genes based on our mapping results. Among these 55 genes, the majority have unsurprisingly well-documented relationships with severe diseases including HIV-1, seizures from epilepsy, leukemia, and Basal Cell Nevus Syndrome, among others. Further, almost all of these genes have documented involvement in cancer development and progression. Table 2 provides a list of four interesting genes to demonstrate the correlation among CpG sites, genes and diseases; the full list of 55 genes can be found as supplementary data (refer to Supplement).

| CpG ID | Associated Gene | Gene Important/Function |
|---|---|---|
| cg23412136 | *MIR548N* | Non-coding region (microRNAs), involved in post-transcriptional regulation of gene expression. |
| cg09720012 | *PAPSS1* | Diseases associated with PAPSS1 include Chronic Monocytic Leukemia. |
| cg17964016 | *DEPTOR* | Inhibits the p53 kinase activity. |
| cg16824024 | *PPP4R2* | Related to pathways of DNA double-strand break repair. |

**Table 2 Relation among CpG sites, associated genes and gene functions.**

Gene MIR548N (Table 2) is a pseudogene for microRNA (mRNA) that controlled by DNA methylation mechanism. This gene product, non-coding RNA molecule, inhibits and destabilizes a target mRNA; the consequences could include down-regulation of the relative gene expression. This gene has been reported as a potential biomarker and therapeutic target for nasopharyngeal carcinoma (Lee et al. 2016). Similarly, the PAPSS1 gene has been associated with chronic monocytic leukemia, and is reported to play an important role in cancer development (Leung et al. 2016). The DEPTOR gene inhibits p53 kinase activity and also appears to play an evolving role in tumor development and progression (Wang et al.

2012). Last but not least, the PPP4R2 gene is related to DNA repair, and has been implicated in a range of cancers (Bosio et al. 2012; Wu et al. 2016; Herzig and Bullinger 2016).

## DISCUSSION

In this paper, we demonstrated the potential of methylation sites as measured in blood samples to serve as biomarkers for cancer. This approach could be a cheap alternative to many standard tests for cancer monitoring and early detection. One important limitation worth noting is the small sample size. The small sample size combined with a large number of features is a fundamental problem that hinders statistical analysis of health data. This can be illustrated from our results, in which we believe accuracy could be increased with a more balanced dataset and larger sample size. These features of the dataset are one reason we utilized an ensemble approach with 4 different methods for variable screening. The machine learning approaches, especially the neural network, did not perform well; however, when we consider that machine learning shines with big data, it's not surprising it was less effective with this small dataset. We illustrated that when sample size is small, traditional statistical models utilizing SAS® Base can be a better solution than machine learning.

There could be a smaller set of variables that could achieve same level of accuracy; therefore, we hope to apply other methods and further reduce the number of variables in future studies. One limitation of this study is the small sample size, i.e. the variable dimension is far larger than sample size. In this study, there were more than 450,000 variables, and the total sample size was only 92. To facilitate our future analysis, we would like to obtain a larger sample size. With more observations, the above-mentioned variable screening method and the follow up machine learning approaches will be more powerful to detect active predictors, and to better forecast the grouping of observations. In that way, our method should be able to differentiate specific cancer types within pan-cancer results, such as identifying breast, colon, or skin cancer in an individual.

## CONCLUSION

In conclusion, we provide the first predictive model for pan-cancer using blood sample methylation profiling data. In this study, we identified 55 CpG sites and the associated genes that have the potential to serve as cancer biomarkers. Using profiling data from GSE89093 of methylation of 55 CpG sites, we built a classification model via linear discriminant analysis. This model allowed us to differentiate between cancer status (healthy/cancer) with high accuracy and low FNR. Other classification algorithms, such as generalized linear regression, SVM, random forest, and neural network, failed to perform well in the face of this ultra-high-dimensional problem with a small dataset.

**ACKNOWLEDGMENTS**

The authors would like to thank SAS and SAS Institute for providing free analytical software, and MWSUG for providing opportunity for this publication.

The authors would also like to thank the anonymous referees for their valuable comments and helpful suggestions.

**CONTACT INFORMATION**

Your comments and questions are valued and encouraged. Contact the author at:

Xi Chen
Department of Molecular and Cellular Biology
University of Kentucky
857-242-1588
billchenxi@gmail.com
www.uky.edu

Jin Xie
Department of Statistics
University of Kentucky
859-684-2081
jin.xie@uky.edu

www.uky.edu

Qingcon Yuan
Assistant Professor
Department of Statistics
Miami University
859-494-3858
qingcong.yuan@miamioh.edu
www.miamioh.edu

## SUPPLEMENT:

| CpG ID | Associated Gene | Gene Important/Function |
|---|---|---|
| cg21330976 | ANKRD29 | Protein coding gene, an analog of this gene is associated with Urethral Diverticulum and Rectal Prolapse |
| cg23412136 | MIR548N | Non-coding region (microRNAs), involved in post-transcriptional regulation of gene expression. |
| cg17383178 | ZBED8 | Protein coding gene. Related to nucleic acid binding. |
| cg24382417 | LOC100499489 | RNA Gene, and is affiliated with the lncRNA class. |
| cg20962444 | DUSP22 | Diseases associated include Alk-Negative Anaplastic Large Cell Lymphoma and Anaplastic Large Cell Lymphoma. |
| cg23985331 | ZFP36L2 | Most likely functions in regulating the response to growth factors. |
| cg02476348 | EOMES | Function as transcription factor which is crucial for embryonic development of mesoderm and the central nervous system in vertebrates. |
| cg13260133 | BORCS7-ASMT | Epresents naturally occurring read-through transcription between the neighboring C10orf32 and AS3MT genes. |
| cg03825236 | RALGAPB | Telated pathways are Vesicle-mediated transport and Translocation of GLUT4 to the plasma membrane. |
| cg02583183 | CFDP1 | May play a role during embryogenesis. |
| cg03052760 | ZMYM4 | Diseases associated with ZMYM4 include Exudative Vitreoretinopathy 1 and Non-Syndromic X-Linked Intellectual Disability. |
| cg13734860 | EIF3E | Among its related pathways are p70S6K Signaling and Viral mRNA Translation. |
| cg12733907 | APPL1 | Diseases associated with APPL1 include Maturity-Onset Diabetes Of The Young, Type 14 and Maturity-Onset Diabetes Of The Young. |
| cg10362527 | GPR1 | May play a role for this receptor in the regulation of inflammation. Can act as a coreceptor for HIV-1. |
| cg13560068 | ZNF789 | Function as *sequence-specific DNA binding*. |
| cg02083376 | EXT2 | Diseases associated with EXT2 include Exostoses, Multiple, Type 2 and Seizures, Scoliosis, And Macrocephaly Syndrome. Among its related pathways are heparan sulfate. |
| cg11287292 | NEIL3 | Function as initiate the first step in base excision repairing (DNA repairing) |
| cg09720012 | PAPSS1 | Diseases associated with PAPSS1 include Chronic Monocytic Leukemia. |
| cg20984590 | KANSL2 | May play a role in chromatin organization. |
| cg17964016 | DEPTOR | Inhibits the p53 kinase activity of both complexes. |
| cg16824024 | PPP4R2 | Related pathways are DNA Double-Strand Break Repair. |
| cg02741329 | LSAMP | The encoded protein may also function as a tumor suppressor and may play a role in neuropsychiatric disorders. |
| cg26103797 | SAP18 | Diseases associated with SAP18 include Basal Cell Nevus Syndrome. |
| cg04774139 | AVPR1A | Diseases associated with AVPR1A include Acth-Independent Macronodular Adrenal Hyperplasia and Asperger Syndrome. |
| cg00046623 | PXDNL | Diseases associated with PXDNL include Hypertrophy Of Breast. |
| cg23978473 | SLC22A23 | *Involves in transmembrane transporter activity*. |
| cg22467473 | RHOBTB3 | Involved in transport vesicle docking at the Golgi complex. |
| cg25925896 | RAB11B | Diseases associated with RAB11B include Rectum Adenocarcinoma and Cystic Fibrosis. |
| cg11752893 | TAP2 | Diseases associated with TAP2 include and . |
| cg06216090 | KLF3-AS1 | An RNA Gene, and is affiliated with the non-coding RNA class. |
| cg27559893 | PRRC2A | Diseases associated with PRRC2A include and . |
| cg00257769 | PARP4 | Related pathways are and . |
| cg12176856 | PDE8A | May be involved in maintaining basal levels of the cyclic nucleotide and/or in the cAMP regulation of germ cell development. |
| cg25513659 | RPH3AL | Rab GTPase effector involved in the late steps of regulated exocytosis, both in endocrine and exocrine cells (By similarity). Diseases associated with RPH3AL include . |
| cg15416250 | CEP128 | |
| cg18363143 | LINC01588 | RNA Gene, and is affiliated with the non-coding RNA class. |
| cg16859636 | PRR26 | RNA Gene, and is affiliated with the ncRNA class. |
| cg12799677 | ALDH9A1 | Diseases associated with ALDH9A1 include and . Among its related pathways are and . |
| cg27482690 | TNRC18 | This gene involves *chromatin binding* and *transcription regulatory region sequence-specific DNA binding*. |
| cg24926589 | CERKL | Diseases associated with CERKL include and . |
| cg06254123 | NDE1 | Diseases associated with NDUFS7 include and . |
| cg00774300 | TSFM | Diseases associated with TSFM include and . |
| cg01358940 | LINC01164 | RNA Gene, and is affiliated with the non-coding RNA class. |
| cg04825119 | AQP6 | The protein encoded by this gene is an aquaporin protein, which functions as a water channel in cells. |
| cg13384284 | LINC00693 | This gene encodes a member of the neuronal calcium sensor (NCS) family of calcium-binding proteins. |
| cg27637873 | NCALD | This gene encodes a member of the neuronal calcium sensor (NCS) family of calcium-binding proteins. |
| cg00188298 | ENTPD1-AS1 | RNA Gene, and is affiliated with the non-coding RNA class. |
| cg15080870 | CCDC9 | Involves *poly(A) RNA binding*. |
| cg10083046 | UBE4A | The encoded protein is involved in multiubiquitin chain assembly and plays a critical role in chromosome condensation and separation through the polyubiquitination of securin. Autoantibodies against the encoded protein may be markers for scleroderma and Crohn's disease. |
| cg10500503 | PLPP3 | Diseases associated with PLPP3 include . |
| cg22113065 | COL4A3BP | Diseases associated with COL4A3BP include and . |
| cg15592324 | TAF4B | Diseases associated with TAF4B include . Among its related pathways are and . |
| cg09726509 | SLC43A1 | Diseases associated with SLC43A1 include . |
| cg27351449 | AHI1 | Diseases associated with AHI1 include and . |
| cg11598353 | NDUFS7 | Diseases associated with NDUFS7 include and . |